%% file: generalized.tex
\documentclass[final]{acmsiggraph}

\title{Boolean Operations using Generalized Winding Numbers}

\graphicspath{{./}{figs/}}

\author{
Alec Jacobson $\qquad$
Columbia University}

\pdfauthor{Alec Jacobson}

\keywords{Constructive solid geometry, boolean set operations, winding number}

\input{pre}
\renewcommand{\X}{\mathcal{X}}
\newcommand{\MA}{A}
\newcommand{\MX}{X}
\newcommand{\MB}{B}
\usepackage{braket}

\begin{document}



\maketitle


\begin{abstract}
The generalized winding number function measures \emph{insideness} for
arbitrary oriented triangle meshes.
Exploiting this, I similarly generalize binary boolean operations to act on
such meshes.
The resulting operations for union, intersection, difference, etc.\ avoid
volumetric discretization or pre-processing.
\end{abstract}

\section{Booleans \& Classic Winding Numbers}
%
%
If $\A⊂\R^3$ and $\B⊂\R^3$ are open subregions of space, then their union
contains all points in $\A$ \emph{or} $\B$, their intersection all points in
$\A$ \emph{and} $\B$, and the difference of $\B$ from $\A$ all points in $\A$
\emph{but not} $\B$. Written in set notation, we have respectively:
\begin{align}
\label{equ:union}
\A ∪ \B         &= \Set{\p | \p ∈ \A \text{ or } \p ∈ \B}, \\
\label{equ:intersect}
\A ∩ \B         &= \Set{\p | \p ∈ \A \text{ and } \p ∈ \B}, \\
\label{equ:difference}
\A \setminus \B &= \Set{\p | \p ∈ \A \text{ and } \p ∉ \B}.
\end{align}

Meanwhile, the winding number function
$w_\A \colon \R^3 \setminus ∂A →\Z$
determines for every point
whether it is inside the set $\A$ purely by examining the set's oriented
boundary $∂\A$.
The winding number integrates the signed surface area of $∂\A$ projected onto a
unit ball around a given point $\p$, or in polar coordinates and w.l.o.g.\
$\p=\mathbf{0}$:
\begin{equation}
w_\A(\p) = \iint\limits_{∂\A} \sin{φ} \,dθ\,dφ, = \begin{cases}
1 & \text{ if $\p ∈ \A$}, \\
0 & \text{ otherwise ($\p ∉ \A$)}.
\end{cases}
\end{equation}
If $\A$ is an (embedded) solid, its winding number will be exactly one for
points inside $\A$ and exactly zero for points outside.

We can replace the set inclusions in the definitions of the boolean operations
in Equations~(\ref{equ:union}-\ref{equ:difference}) with winding number
expressions:
replace \mbox{$\p ∈ \X$} with $w_\X(\p) = 1$ and $\p ∉ \X$ with $w_\X(\p) =0$.

\begin{figure}[t]
\includegraphics[width=\linewidth]{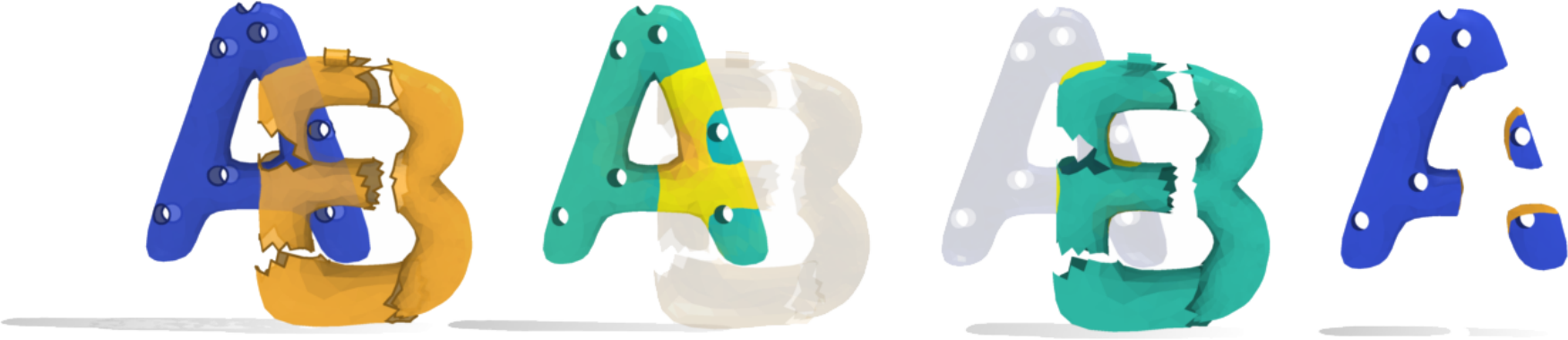}
\caption{Holey, non-manifold triangle meshes $\MA$ and $\MB$ each induce a
generalized winding number function on the other's faces, useful for boolean
operations such as $\MA \setminus \MB$.}
\label{fig:teaser}
\end{figure}

\paragraph{Algorithm}
This immediately reveals an algorithm for conducting boolean operations on
\emph{solid}%
\footnote{A solid triangle mesh is the oriented boundary of solid region: its
induced winding number is exactly zero or one for any point not on the mesh.}
triangle meshes $\MA$ and $\MB$
with vertices at general positions in space $\MA$ and $\MB$.
Refine each mesh at mutual triangle-triangle intersections with the other so
that intersections lie only at vertices and along edges (see, e.g.,
\cite{Jacobson:WN:2013}).
For each triangle $t_a$ of $\MA$, determine its winding number with respect to
$\MB$, e.g., by evaluating $w_\MB(\t_a)$ where $\t_a$ is the triangle $t_a$'s
barycenter.
Likewise determine for each triangle $t_b$ of $\MB$ its winding number with
respect to $\MA$.
Finally---depending on the boolean operation---keep, keep-and-flip, or discard
each triangle from $\MA$ and $\MB$.
For example, for $\A \setminus \B$ (see \reffig{teaser}),
\begin{align}
\forall\,t_a ∈ \MA & \begin{cases}
  \text{keep\hphantom{-and-flip}}    & \text{ if $w_\MB(\t_a) = 0$}, \\
  \text{discard}                     & \text{ otherwise $\left(w_\MB(\t_a) = 1\right)$},
\end{cases}\\
\forall\,t_b ∈ \MB & \begin{cases}
  \text{keep-and-flip}               & \text{ if $w_\MA(\t_b) = 1$}, \\
  \text{discard}                     & \text{ otherwise $\left(w_\MA(\t_b) = 0\right)$}.
\end{cases}
\end{align}

For non-general position meshes, if the barycenter $\t_a$ from a triangle of
$\MA$ lies exactly on $\MB$ (e.g., due to coplanar overlaps), 
the winding number $w_\MB(\t_a)$ is undefined and this algorithm will fail.
Coplanar overlaps can be handled by identifying perfectly coplanar triangles
during triangle-triangle intersection resolution.
For each pair of resolved ``duplicate'' triangles $t_a$ and $t_b$, either keep
one or discard both depending on the operation.
For example, if conducting the difference of $\B$ from $\A$, then discard both
if $t_a$ and $t_b$ have the same orientation and keep $t_a$ otherwise.

The classic notion of the winding number applies to any closed oriented
surface, not just those bounding a solid.
The winding number for such surfaces will be a possibly negative integer
indicating \emph{how many times} a point is inside the surface.
We can immediately generalize our algorithm to handle any immersed closed
oriented triangle meshes by replacing $w_\MX(\p) = 1$ with $w_\MX(\p) > 0$.
Alternatively, if we consider ``inside-out'' regions as inside then replace
$w_\MX(\p) = 1$ with $|w_\MX(\p)| > 0$.


\section{Generalized Winding Numbers}
Jacobson et al.\ 
\shortcite{Jacobson:WN:2013}
define the generalized winding number $w_\MX \colon \R^3\setminus
\MX→\R$ for arbitrary oriented triangle meshes as the sum of the signed solid
angles $Ω_t$ subtended by each triangle $t$:
\begin{equation}
w_\MX(\p) = ∑\limits_{t ∈ \MX} \frac{1}{4π} Ω_t(\p).
\end{equation}
This smooth function measures for each point in space \emph{how much it is
inside} $\X$.
For example, a punctured sphere will induce a winding number value close to one
inside and far from the hole, and a value close to zero outside and far from
the hole.
Near the hole, the value will smoothly transition from near one to near zero
passing through one half close by the hole.

We may immediately generalize the boolean algorithm in the previous section to
arbitrary oriented triangle meshes.
Simply replace  $w_\MX(\p) = 1$ with $w_\MX(\p) > 1/2$;
alternatively, with $|w_\MX(\p)| > 1/2$ if treating negative regions as inside.
%

%
This proposed method avoids volumetric discretization of Jacobson et al.\
\shortcite{Jacobson:WN:2013} but enjoys their hierarchical winding number
evaluation.
%
Computing the generalized winding number as a floating point summation could
potentially lead to incorrect assignment due to round-off.
%
If evaluation points (i.e., barycenters from the \emph{other} mesh) are not
very close to open boundaries then this is a non-issue.

\renewcommand\refname{Reference}
\bibliographystyle{acmsiggraph}
\bibliography{references}

\end{document}

%% file: pre.tex
\usepackage[table]{xcolor}
\usepackage{tabularx}
\usepackage{booktabs}
\usepackage{pdfcomment}
\usepackage{amsmath}

\usepackage{amsfonts}
\usepackage{amssymb}

\usepackage[english]{babel}

\definecolor{lightbluishgrey}{rgb}{0.78,0.86,0.93}

\definecolor{highlightgreen}{rgb}{0.1,0.7,0.1}

\newcommand{\reviewer}[1]{{}} 

\newcommand{\LK}[1]{}

\newcommand{\cheatvspace}[1]{}

\newcommand{\PWN}{}
\def\PWN/{PWN}
\newcommand{\numpwn}{}
\def\numpwn/{8616}

\usepackage{layouts}

\usepackage{wrapfig}

\newcommand{\reffig}[1] {Figure~\ref{fig:#1}}

\usepackage{xfrac}

\usepackage{amsmath}
\usepackage{amssymb}    
\usepackage{cancel}
\usepackage[T1]{fontenc}

\newcommand{\R}{\mathbb{R}}

\newcommand{\vc}[1]{\mathbf{#1}}

\newcommand{\p}{\vc{p}}

\renewcommand{\t}{\vc{t}}

\newcommand{\A}{\mathcal{A}}
\newcommand{\B}{\mathcal{B}}

\newcommand{\X}{\mat{X}}

\newcommand{\Z}{\mathbb{Z}}

\usepackage{siunitx}
\usepackage[mathletters]{ucs}
\usepackage[utf8x]{inputenc}
\usepackage{upgreek}
\usepackage{dblfloatfix}

\usepackage{graphicx}

\usepackage{wrapfig}

\makeatletter
\let\save@mathaccent\mathaccent
\newcommand*\if@single[3]{%
  \setbox0\hbox{${\mathaccent"0362{#1}}^H$}%
  \setbox2\hbox{${\mathaccent"0362{\kern0pt#1}}^H$}%
  \ifdim\ht0=\ht2 #3\else #2\fi
  }
\newcommand*\rel@kern[1]{\kern#1\dimexpr\macc@kerna}
\newcommand*\widebar[1]{\@ifnextchar^{{\wide@bar{#1}{0}}}{\wide@bar{#1}{1}}}
\newcommand*\wide@bar[2]{\if@single{#1}{\wide@bar@{#1}{#2}{1}}{\wide@bar@{#1}{#2}{2}}}
\newcommand*\wide@bar@[3]{%
  \begingroup
  \def\mathaccent##1##2{%
    \let\mathaccent\save@mathaccent
    \if#32 \let\macc@nucleus\first@char \fi
    \setbox\z@\hbox{$\macc@style{\macc@nucleus}_{}$}%
    \setbox\tw@\hbox{$\macc@style{\macc@nucleus}{}_{}$}%
    \dimen@\wd\tw@
    \advance\dimen@-\wd\z@
    \divide\dimen@ 3
    \@tempdima\wd\tw@
    \advance\@tempdima-\scriptspace
    \divide\@tempdima 10
    \advance\dimen@-\@tempdima
    \ifdim\dimen@>\z@ \dimen@0pt\fi
    \rel@kern{0.6}\kern-\dimen@
    \if#31
      \overline{\rel@kern{-0.6}\kern\dimen@\macc@nucleus\rel@kern{0.4}\kern\dimen@}%
      \advance\dimen@0.4\dimexpr\macc@kerna
      \let\final@kern#2%
      \ifdim\dimen@<\z@ \let\final@kern1\fi
      \if\final@kern1 \kern-\dimen@\fi
    \else
      \overline{\rel@kern{-0.6}\kern\dimen@#1}%
    \fi
  }%
  \macc@depth\@ne
  \let\math@bgroup\@empty \let\math@egroup\macc@set@skewchar
  \mathsurround\z@ \frozen@everymath{\mathgroup\macc@group\relax}%
  \macc@set@skewchar\relax
  \let\mathaccentV\macc@nested@a
  \if#31
    \macc@nested@a\relax111{#1}%
  \else
    \def\gobble@till@marker##1\endmarker{}%
    \futurelet\first@char\gobble@till@marker#1\endmarker
    \ifcat\noexpand\first@char A\else
      \def\first@char{}%
    \fi
    \macc@nested@a\relax111{\first@char}%
  \fi
  \endgroup
}
\makeatother

\usepackage[ruled,vlined]{algorithm2e}
\usepackage{etoolbox}

  \makeatletter
  \patchcmd{\algocf@Vline}{\vrule}{\vrule\hspace{-0.25em}}{}{}
  \makeatother
  \DontPrintSemicolon
  \SetAlgoLined
  \SetKwInput{KwData}{Inputs} 
  \SetKwInput{KwResult}{Outputs}
  \RestyleAlgo{ruled}
  \SetKwBlock{Repeat}{repeat}{}

\usepackage{array}

\makeatletter
\newcommand{\raisemath}[1]{\mathpalette{\raisem@th{#1}}}
\newcommand{\raisem@th}[3]{\raisebox{#1}{$#2#3$}}
\makeatother

\newcommand{\thetitle}{}
\def\thetitle/{Mesh Arrangements for Solid Geometry}